\documentclass[runningheads]{llncs}
\usepackage{graphicx}
\usepackage{amsmath}
\usepackage[labelfont=bf]{caption}                         
\usepackage[subrefformat=parens]{subfig}
\usepackage{enumitem}  
\usepackage{amssymb}
\usepackage[misc]{ifsym}
\usepackage{graphicx}
\usepackage{threeparttable}
\usepackage{ntheorem}
\usepackage{color}
\usepackage{colortbl}
\usepackage{multirow}
\usepackage{amsmath}
\usepackage{comment}
\usepackage{bm}                                            
\usepackage{etoolbox}                                      
\usepackage{url}
\usepackage{nth}
\usepackage{cite}
\usepackage{balance}
\usepackage[bookmarks=false]{hyperref}                     %
\usepackage{courier}                                       
\usepackage{listings}                                      

\begin{document}

\title{NeRD: Neural Representation of Distribution for Medical Image Segmentation}




\authorrunning{H. Zhang et al.}

\institute{Cornell University \\ \email{hz459@cornell.edu} \and
Weill Cornell Medical College \and
University of Pennsylvania
}

\author{Hang Zhang\inst{1,2} \Letter \and
Rongguang Wang \inst{3} \and
Jinwei Zhang \inst{1,2} \and
Chao Li \inst{1,2} \and
Gufeng Yang \inst{1} \and
Pascal Spincemaille \inst{2} \and
Thanh D. Nguyen \inst{2} \and
Yi Wang \inst{1,2}
}



\maketitle

\begin{abstract}

We introduce Neural Representation of Distribution (NeRD) technique, a module for convolutional neural networks (CNNs) that can estimate the feature distribution by optimizing an underlying function mapping image coordinates to the feature distribution. 
Using NeRD, we propose an end-to-end deep learning model for medical image segmentation that can compensate the negative impact of feature distribution shifting issue caused by commonly used network operations such as padding and pooling.
An implicit function is used to represent the parameter space of the feature distribution by querying the image coordinate.
With NeRD, the impact of issues such as over-segmenting and missing have been reduced, and experimental results on the challenging white matter lesion segmentation and left atrial segmentation verify the effectiveness of the proposed method. The code is available via https://github.com/tinymilky/NeRD.


\end{abstract}

\section{Introduction}

Deep convolutional neural networks (CNNs) have been the dominant approach across various tasks of computer vision and image processing.
The efficient convolutional operation with spatially-invariant filters is one of the main factors for the success of CNNs. 
The input feature map to the convolutional layers shares these filters across all spatial positions, thereby reducing network parameters and improving the generalization ability.
Many medical image applications have benefited from this, for example, multiple sclerosis (MS) lesion segmentation \cite{zhang2020efficient,zhang2020geometric}, white matter (WM) lesion segmentation \cite{la2020multiple}, and quantitative susceptibility mapping \cite{zhang2020fidelity,zhang2020bayesian}.

\begin{figure}[!t]
	\centering
	\includegraphics[width=.6\columnwidth,height=0.3191\columnwidth]{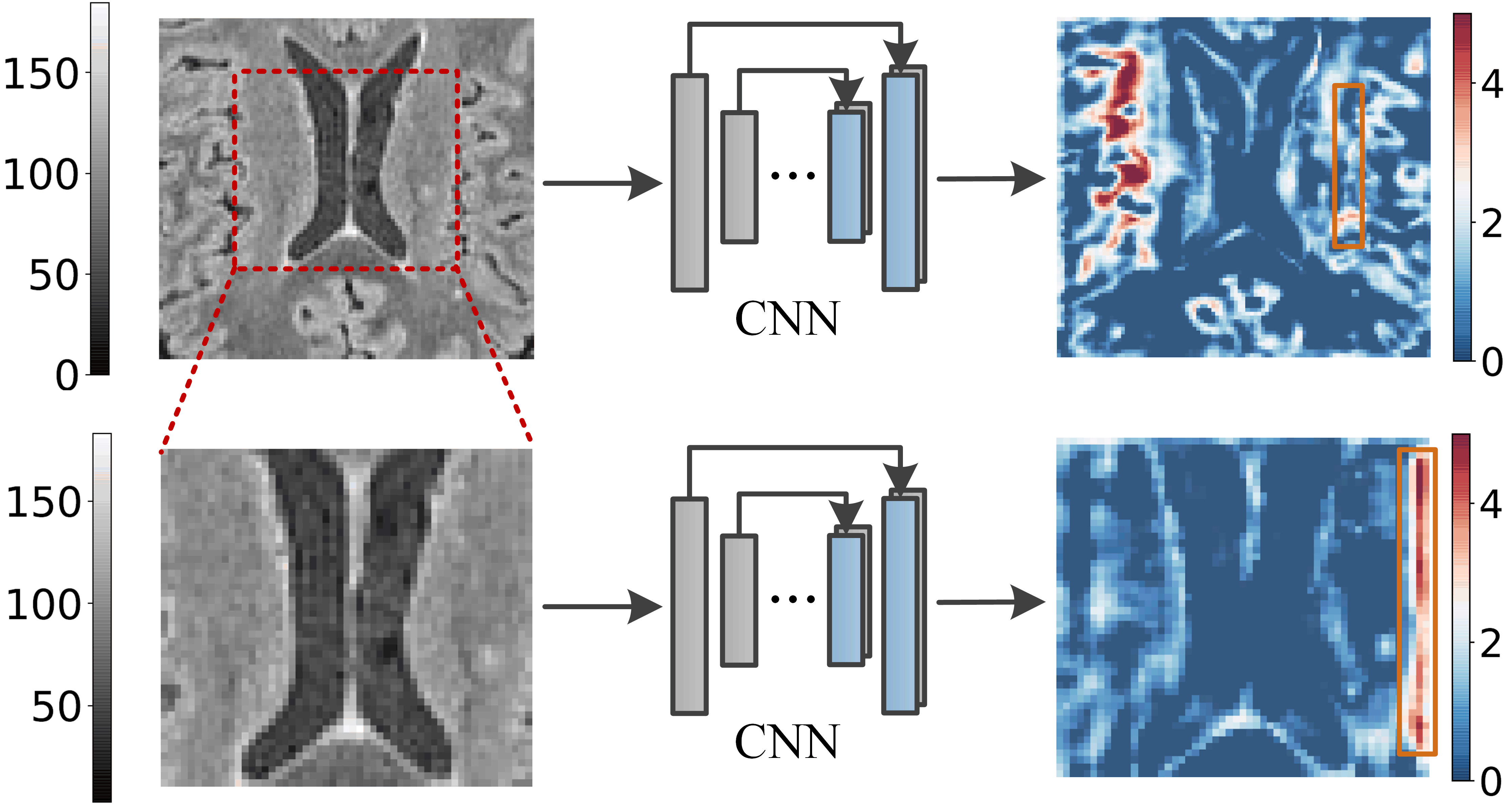}
	\caption{
	Visual example of the feature shifting. 
	The image at the left lower panel is cropped from the image at the left upper panel.
	The feature maps on the right of both images are obtained with a U-Net.
	Orange boxes indicate regions of feature shifting.
	}
	\label{fig:artefact}
\end{figure}

However, we still observe severe failure cases when applying the widely used U-Net \cite{ronneberger2015u} to brain lesion segmentation.
E.g., We can see from Fig.~\ref{fig:wmh_example} and Fig.~\ref{fig:wmh_example} that brain lesions close to the brain boundary (meaning that they are close to the image boundary) or close to the ventricles (meaning that they are close to the center of the image) are prone to be misclassified by the U-Net.
Analyzing the network architecture, we summarize the causes as follows: 1) the padding operation in convolutional layers can lead to artefacts in feature maps \cite{liu2008reducing,alsallakh2020mind} (see Fig.~\ref{fig:artefact}), 2) and can shift the feature distribution across different spatial positions \cite{islam2019much}; 3) the down-sampling operations such as max-pooling and strided convolution ignore the basic sampling theorem, resulting in breaking the property of spatial invariance for the segmentation task \cite{zhang2019making,kayhan2020translation}.
\begin{figure}[!t]
	\centering
    \includegraphics[width=1.0\columnwidth,height=0.2964\columnwidth]{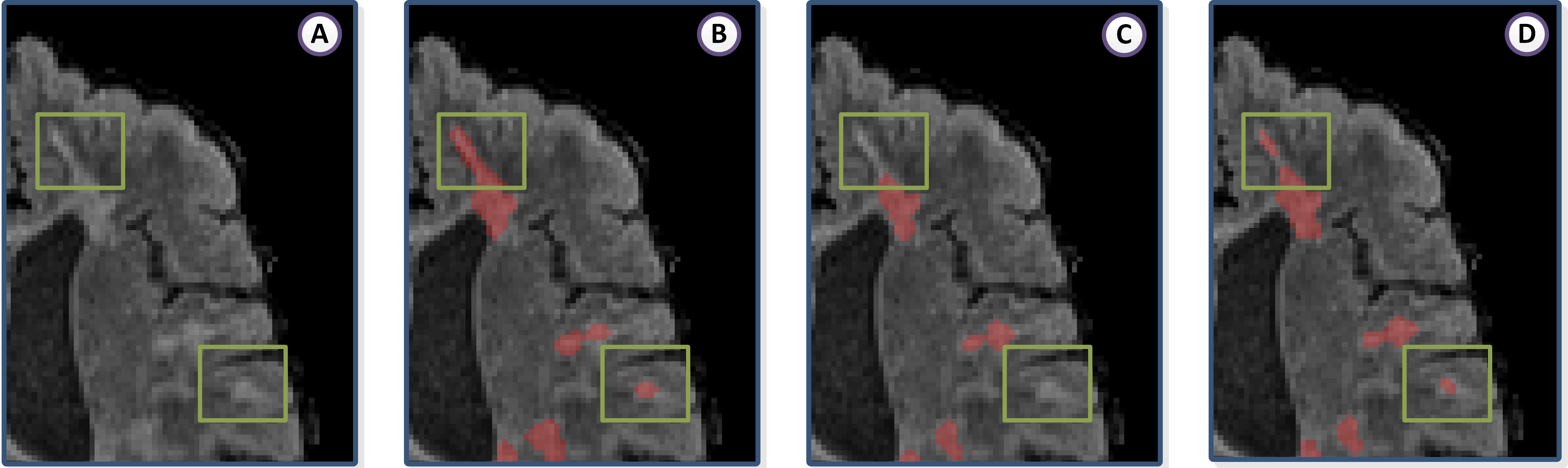}
	\caption{ 
	Visualization of a failure case on brain lesion segmentation. 
	Green boxes indicate regions of failure.   
	(A) An T2-FLAIR image example of a patient with heavy lesion burden.
	(B) Lesions labeled by a human expert (marked in red). 
	(C) Segmentation result of U-Net. 
	(D) Segmentation result of our proposed CIF-based network.
	}
	\label{fig:failure_case}
\end{figure}
Various methods, such as cube padding \cite{cheng2018cube}, circular convolution \cite{schubert2019circular}, explicit boundary-based filters \cite{innamorati2019learning}, and the max-blurring-pooling \cite{zhang2019making} have been investigated by researchers to tackle the above issues.
All of these methods can lessen the negative impact in some degree, but most of these methods are ad-hoc and none of them can handle all three issues collectively.
Therefore, we argue that it is imperative to develop a unified framework to solve the problem to facilitate our medical image segmentation.

The key problem is the feature distribution shifting caused by commonly used network operations such as padding and pooling.
Usually, the deep network stacks many convolutional layers; thus, with the network depth increases, the consecutive operations of padding and pooling can gradually shift the feature distribution from the boundary to the center of the image.
Suppose $\Omega \subset \mathbb{N}^{2}$ is the spatial domain of an input image, $\mathbf{v}=(i, j) \in \Omega$ is the spatial position vector, the final feature map before the pixel-wise classifier is $\mathbf{X} \in \mathbb{R}^{H \times W\times C}$, $\mathbf{x}_{\mathbf{v}} \in \mathbb{R}^{C}$ is the feature vector in position $\mathbf{v}$, and $\phi(\theta)$ is the feature vector distribution, then we can define the following mapping function:
\begin{align}
    \theta = f(\mathbf{v}),  \label{eq:theat_f_v} \\ 
    \mathbf{X}_{\mathbf{v}} \sim \phi(\theta),
\end{align}
where $\theta$ is the parameter of the distribution $\phi$.
Basically, the equation tells us the distribution of the feature is determined by its spatial location.
If convolutional neural network is strictly spatially invariant, the $\theta$ should be equivalent for features in all locations.
However, as we mentioned above, certain operations in the network can alter the distribution, resulting in varying $\theta$ depending on its location. 

Current segmentation network assumes that $\theta$ is equivalent for all feature vectors, which brings trouble for our brain lesion segmentation.
Thus, in this paper, we propose a Neural Representation of Distribution (NeRD) technique to approximate the mapping function $f$ in Eq.~\eqref{eq:theat_f_v} to resolve the issue.
The idea of NeRD is inspired by a recently developed neural implicit representation (NIR) technique \cite{park2019deepsdf,mildenhall2020nerf}, where the NIR parameterizes a signal as a continuous function that maps the domain such as the coordinate of the signal to pixel values at that coordinate.
In our study, we map the coordinates to the feature distributions.
In summary:
\begin{itemize}
    \item We propose a Neural Representation of Distribution (NeRD) technique to bring back the spatial invariance by approximating the feature distribution based on pixel coordinates.
    \item We validate the proposed method on two challenging medical image segmentation tasks, where both quantitative and qualitative results demonstrate the effectiveness of our method.
\end{itemize}

\subsection{Related Works}

\subsubsection{Brain Lesion Segmentation}

MS lesion segmentation and WM lesion segmentation are most important and difficult tasks in brain lesion analysis, as these lesions vary greatly in terms of shape, size and location. 
Though numerous automated approaches have been proposed, a clinically reliable technique is not yet available.
The 2.5D stacked denseNet \cite{zhang2019multiple} is proposed to capture broader brain structure information.
The folded attention network \cite{zhang2020efficient} applies light-weight self-attention method for richer contextual information.
The geometric loss \cite{zhang2020geometric} is developed to regularize CNN training, which helps segmenting small lesions.
The boundary loss \cite{kervadec2019boundary} uses distance transformation mapping to tackle the data imbalance problem.
All these methods have achieved reasonably good result, but bone of them takes feature distribution shifting into consideration.

\subsubsection{Neural Implicit Representation}

NIR is a recently developed technique, which is frequently used in representation of 
geometry and appearance \cite{park2019deepsdf,mildenhall2020nerf} in graphics.
DeepSDF \cite{park2019deepsdf} Learns a set of continuous signed distance functions for shape representation.
Later, NeRF \cite{mildenhall2020nerf} provides a more flexible way for synthesizing novel views of complex scenes.
Other vision applications such as image super resolution \cite{chen2020learning}, image synthesis \cite{anokhin2020image} also benefit from the NIR technique.
Though the NIR is blooming in many areas, we haven't seen any method using NIR for feature distribution mapping. 

\subsubsection{Meta Learning}

We use the proposed NeRD technique to predict network weights based on image coordinates, the process of which is one of the meta learning strategies \cite{lemke2015metalearning}.
The weights of certain modules of the network are predicted by another network module instead of directly learned.
\cite{noh2016image} proposes a dynamic parameter layer for image question answering.
\cite{hu2018learning} uses box weights to predict mask weights for image segmentation.
\cite{hu2019meta} achieves super-resolution of arbitrary scale factor by predicting weights of up-sampling module based on the scale factor.
In our work, we use image coordinates coupled with training data to dynamically estimate the feature distribution.
\section{Methodology}

Recently, NIR techniques have been investigated to model continuous 3D shapes as level sets, which can efficiently map coordinates to signed distance function \cite{jiang2020local,park2019deepsdf} for shape reconstruction.
In this work, our goal is to map the image coordinates to the feature distribution to resolve the spatial invariance issue brought by basic operations such as padding and pooling.

\subsection{Neural Representation of Distribution}

The practical implementation of a segmentation network consists of two parts, a encoder-decoder structure for feature extraction and a multi-layer perceptron (MLP) for pixel-wise classification.
Let $\mathbf{X} \in \mathbb{R}^{H\times W \times C}$ ($H$ and $W$ are the spatial size of the image, and $C$ is the number of channels) be the output from the encoder-decoder structure, $\mathbf{v}=(d_t,d_r,d_b,d_l)$ be the position vector for a pixel ($d_t,d_r,d_b$, and $d_l$ indicates the distance of the pixel to the top, the right, the bottom and the left of the image), we can use the Gaussian distribution to approximate the feature distribution of a given position $\mathbf{v}$ as follows:
\begin{equation}
    \mathbf{x}_{\mathbf{v}} \sim  \mathcal{N}(\mathbf{\mu},\mathbf{\Sigma}),
\end{equation}
where $\mathbf{\mu}$ is the mean vector in location $\mathbf{v}$, and $\mathbf{\Sigma}$ is the corresponding co-variance matrix.
Since the spatial invariance of CNNs has been broken by certain operations in the network, one of the MLP assumptions that the data is drawn from the same distribution no longer holds.
It is expected that a simple MLP classifier is prone to fail in classifying pixels close to the boundary or the center of the image, as there exists an non-trivial discrepancy between their feature distributions (as is verified in our experiments, see Fig.~\ref{fig:wmh_example} and Fig.~\ref{fig:atrial_example}).
In this work, we use the proposed NeRD technique to resolve the issue.
\begin{figure}[!t]
	\centering
	\includegraphics[width=1.0\columnwidth,height=0.3286\columnwidth]{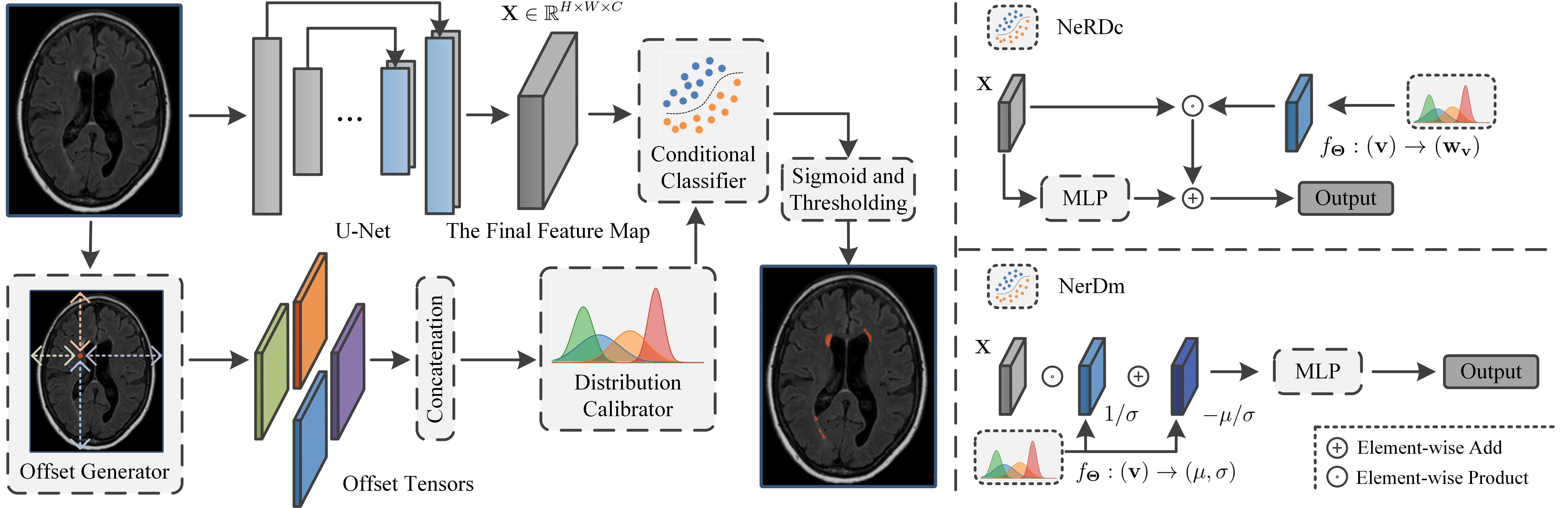}  
	\caption{
	Visual illustration of the proposed method.
	}
	\label{fig:framework}
\end{figure}

We represent a continuous parameter space of the distribution function as a 4D vector-valued function, whose input is a 4D position vector $\mathbf{v}=(d_t,d_r,d_b,d_l)$, and whose output is the mean vector $\mathbf{\mu}$  and the co-variance matrix $\mathbf{\Sigma}$ of the distribution of this position.  
In practice, we use another MLP $f$ to approximate this continuous 4D representation of the distribution, and optimize the weight $\mathbf{\Theta}$ of $f$ along with other network weights to map each input position vector to the corresponding mean vector and co-variance matrix.
It has been shown that CNNs can linearize \cite{bengio2013better,upchurch2017deep} the manifold of images into an Euclidean subspace of deep features, which indicate that elements in the output feature vector are independent to each other. 
Thus, we further simplify the model and reduce the co-variance matrix to a vector $\mathbf{\sigma}$.
With the estimation of $\mathbf{\mu}$ and $\mathbf{\sigma}$ from the function $f$, we can normalize and classify the output feature vector at position $\mathbf{v}$ as follows:
\begin{equation}
s_{\mathbf{v}} = \mathbf{w}^{\top}\dfrac{\mathbf{x}_{\mathbf{v}} - \mathbf{\mu}}{\mathbf{\sigma}},
\end{equation}
where $\mathbf{\mu}$ and $\mathbf{\sigma}$ are obtained with the MLP $f_{\mathbf{\Theta}}:(\mathbf{v})\rightarrow (\mathbf{\mu},\mathbf{\sigma})$, $s_{\mathbf{v}}$ is the final output of the network, $\mathbf{w}$ is the weight of the MLP classifier (Please note that as we use z-score \cite{zill2020advanced} to normalize the input image, there is no bias term for the classifier).
To be numerically stable, in practice, we estimate $1/\mathbf{\sigma}$ and $-\mu/\mathbf{\sigma}$
instead, and the final equation for normalization and classification can be described as:
\begin{equation}
s_{\mathbf{v}} = \mathbf{w}^{\top} (\mathbf{x}_{\mathbf{v}}\dfrac{1}{\mathbf{\sigma}} + (- \dfrac{\mathbf{\mu}}{\mathbf{\sigma}}))
\label{eq:musig}
\end{equation}

\subsection{Pixel-aligned Classifier and The Overall Framework}

The overall framework of our proposed method is shown in Fig.~\ref{fig:framework}.
The input image goes through a U-Net to obtain the final feature map for pixel-wise classification, and in the meantime, the offset generator provides the pixel-wise position vectors, followed by an MLP as the distribution calibrator to generate the estimation of $1/\mathbf{\sigma}$ and $-\mu/\mathbf{\sigma}$.
With the distribution estimation of every pixel position, we apply Eqn.~\eqref{eq:musig} to normalize and classify the feature vector of each position.
The final segmentation can be obtained with another MLP and a Sigmoid function.
We call the framework using the proposed feature estimation technique as NeRD.

We wouldn't say NeRD a concrete method but an idea that can improve the performance of a CNN network by estimating the pixel-wise feature distribution.
Thus, it can be expected that there are many variants using the NeRD idea.
Here, we describe one that we deem as interesting.
Rather than estimating the parameters of the feature distribution, we estimate a pixel-aligned classifier.
That is to say, we estimate a unique linear classifier for every single pixel position, which can be described as $s_{\mathbf{v}} = \mathbf{x}_{\mathbf{v}}^{\top} \mathbf{w}_{\mathbf{v}}$, where the $\mathbf{w}_{\mathbf{v}}$ is the weight of a linear classifier estimated by an MLP $f_{\mathbf{\Theta}}:(\mathbf{v})\rightarrow (\mathbf{w}_{\mathbf{v}})$. 
We call this NeRD classifier as NeRDc and the former mean-variance estimator as NeRDm.
Both NeRDc and NeRDm can be efficiently implemented using tensor operations in the modern GPU architecture.
\section{Experimental Results}

\subsection{Datasets}

White matter hyperintensities\footnote{\url{https://wmh.isi.uu.nl}} (WMH)~\cite{kuijf2019standardized} is a publicly available dataset which contains 60 3D scans with 2 modalities (T1 and FLAIR weighted) acquired from multiple vendors and scanners in three different institutes. 
The spatial resolution goes from $0.95\times0.95\times3\ mm^3$ to $1.21\times1\times3\ mm^3$ for each volume. 
Manual annotations of WMH are provided for the 60 scans.
In the experiments, we split this dataset into training, validation, and testing sets containing 42, 6 and 12 samples, respectively.

We also use left atrial\footnote{\url{http://atriaseg2018.cardiacatlas.org}} (LA) segmentation challenge~\cite{xiong2021global} dataset for evaluation.
A total of 154 independently acquired 3D LGE-MRIs from 60 deidentified patients with atrial fibrillation were used in this challenge.
The clinical images were acquired with either a 1.5T Avanto or 3.0T Verio whole-body scanner.
The spatial resolution of one 3D LGE-MRI scan was $0.625\times0.625\times0.625\ mm^3$ with spatial dimensions of either $576\times576\times88$ or $640\times640\times88$ pixels. 
From the whole set, 108 scans were used for training, 15 for validation, and the remaining 31 for testing.

\subsection{Implementation Details}

\subsubsection{Data pre-processing.} We slice the original images into a stack of independent 2D images. 
Each scan is center-cropped to size $160\times224$ for WMH and size $290\times240$ pixels for LA, and normalized to real values between 0 and 1.
Since two modalities (T1 and FLAIR) are available for WMH, both of them are concatenated along the channel dimension before being used as input to the network. 

\subsubsection{Netowrk and training.} We employ U-Net~\cite{ronneberger2015u} as the backbone architecture in our experiments. 
To train our model, we use Adam~\cite{kingma2014adam} optimizer, with an initial learning rate of $1\text{e}-3$ (weight decay of $1\text{e}-6$) and a batch size equal to 14.
The learning rate is halved at 50\%, 70\% and 90\% of the total training epoch (90) for optimal convergence.
We use PyTorch~\cite{paszke2019pytorch} for implementation, and run the experiments on a machine equipped with an NVIDIA RTX 2080 Ti GPU with 11GBs of memory.

\subsubsection{Evaluation metrics.} To quantify the performance of WMH segmentation, we use Dice similarity coefficient~\cite{dice1945measures}, lesion-wise Dice (LDice), lesion-wise true positive rate (LTPR), and lesion-wise positive pre-dictive value (LPPV) as metrics.
LDice, LTPR and LPPV are defined as $\text{LDice} = \frac{\text{TPR}}{\text{GL}+\text{PL}}$, $\text{LTPR} = \frac{\text{TPR}}{\text{GL}}$, and $\text{LPPV} = \frac{\text{TPR}}{\text{PL}}$, where TPR denotes the number of lesions in ground-truth segmentation that overlap with a lesion in the produced segmentation, and GL, PL is the number of lesions in ground-truth segmentation and produced segmentation respectively.
Dice quantifies the voxel-wise overlap between the output and the ground-truth.
Complementarily, LDice, LTPR and LPPV are more sensitive in measuring the lesion-wise detection accuracy.
As for LA segmentation, two region-based metrics, Dice and Jaccard, are used to measure the region mismatch. 
Three boundary-based metrics, average surface distance (ASD), Hausdorff distance (HD), and 95\% Hausdorff distance (95HD), are used to evaluate errors in the boundary.

\subsection{WMH Segmentation}

\begin{table}[!t]
\caption{Quantitative comparison with average (standard deviation over three independent runs) on white matter hyperintensities (WMH) segmentation.}
\label{tab:wmh_table}
\begin{center}
\setlength{\tabcolsep}{4pt}
\begin{tabular}{ l ccccc }
\hline
\hline
Model & Filter & Dice (\%) $\uparrow$ & LDice (\%) $\uparrow$ & LFPR (\%) $\downarrow$ & LTPR (\%) $\uparrow$ \\
\hline
U-Net & 256 & 76.7 (1.96) & 66.9 (5.37) & 30.8 (4.95) & 73.9 (4.15) \\
NeRDc & 256 & \bf{78.4 (0.41)} & \bf{69.7 (1.84)} & \bf{27.8 (3.52)} & 72.2 (0.94)\\
NeRDm & 256 & 77.4 (0.71) & 66.9 (2.82) & 33.5 (3.71) & \bf{73.9 (0.86)}\\
\hline
U-Net & 512 & 78.4 (0.82) & 69.8 (0.55) & 29.2 (3.14) & 76.2 (3.07)\\
NeRDc & 512 & \bf{79.2 (0.05)} & 72.0 (1.07) & \bf{27.1 (0.44)} & \bf{77.7 (1.83)}\\
NeRDm & 512 & 78.6 (0.37) & \bf{72.3 (0.32)} & 27.7 (1.11) & 77.6 (2.43)\\
\hline
\hline
\end{tabular}
\end{center}
\end{table}

\begin{figure}[!t]
\centering
\includegraphics[width=1.0\columnwidth,height=0.2786\columnwidth]{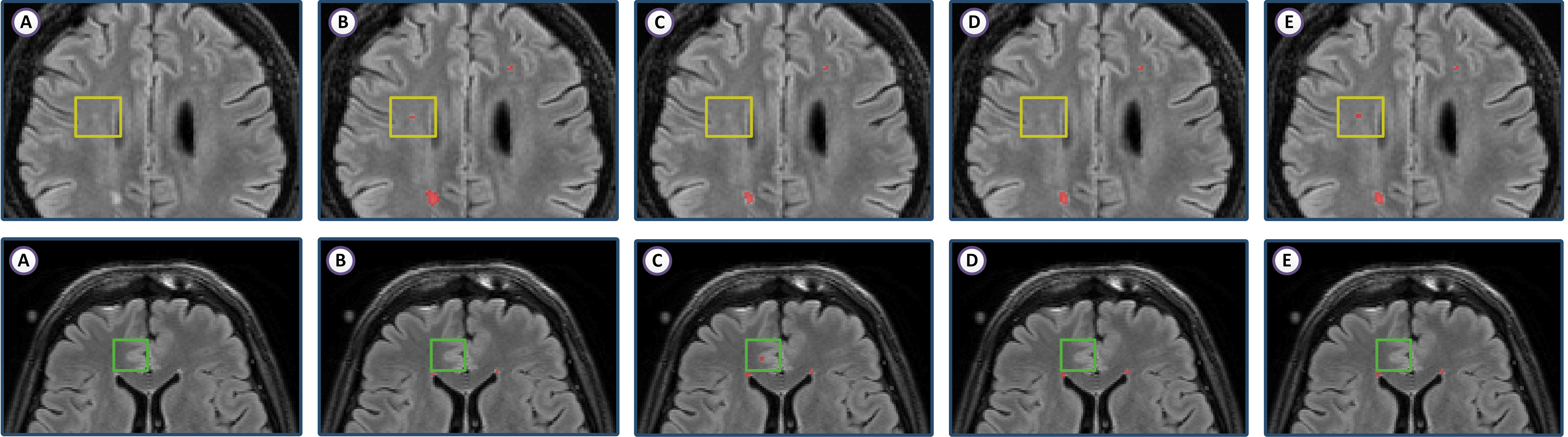}  
\caption{
WMH examples. 
Yellow boxes indicate regions of missing and green boxes indicate regions of over-segmenting. (A) FLAIR images; (B) ground-truth (marked in red); (C) U-Net mask; (D) NeRDc mask; (E) NeRDm mask.
}
\label{fig:wmh_example}
\end{figure}

\subsubsection{Quantitative results.}
We compare our proposed NeRDc and NeRDm methods with the baseline U-Net~\cite{ronneberger2015u} on lesion-specific metrics as shown in Table~\ref{tab:wmh_table}. 
We use two variants of U-Net backbone containing different set of kernel channel numbers that representing low- and high-capacity networks separately. 
For the low-capacity network (denoted as 256 under ``filter'' column in the table), we use [16, 32, 64, 128, 256] as the size for convolution filters in different layers, and we double the channel number for the high-capacity network (denoted as 512).
In the low-capacity group, we observe that NeRDc outperforms both U-Net and NeRDm in most metrics, and in particular, there's 2.8\% increment for LDice and 3\% reduction in LFPR compared to U-Net.
NeRDc also shows substantial improvements over Dice (0.8\%), LFPR (2.1\%), and LTPR (1.5\%) in the high-capacity group compared to U-Net.
Importantly, our proposed NeRDc using low-capacity backbone achieved similar performance compared to the U-Net with high-capacity backbone.
For example, the Dice score of NeRDc in low-capacity group is the same as the U-Net in high-capacity group.

\subsubsection{Qualitative results.}
As shown in Fig.~\ref{fig:wmh_example}, we present WMH segmentation results with a missing case and a over-segmenting case made by vanilla U-Net.
We can see from the first row of Fig.~\ref{fig:wmh_example}, U-Net missed a lesion close to the ventricle/center of the image, while our NeRDm accurately located this lesion, which demonstrated the superiority of our feature distribution estimation technique.
Though the end of cortex (yellow box position) exhibited similar high intensity value as lesions, and was close to the boundary, both our NeRDc and NeRDm made no mistakes in this region, while U-Net over-segmented several pixels.

\subsection{LA Segmentation}

\begin{table}[!t]
\caption{Quantitative comparison with average (standard deviation over three independent runs) on left atrial (LA) segmentation.}
\label{tab:la_table}
\begin{center}
\setlength{\tabcolsep}{2.5pt}
\begin{tabular}{ l cccccc }
\hline
\hline
Model & Filter & Dice (\%) $\uparrow$ & Jaccard (\%) $\uparrow$ & HD $\downarrow$ & 95HD $\downarrow$ & ASD $\downarrow$ \\
\hline
U-Net & 256 & 90.3 (0.51) & 82.5 (0.81) & 34.1 (3.09) & 6.5 (0.52) & 2.1 (0.13) \\
NeRDc & 256 & \bf{90.6 (0.26)} & \bf{82.9 (0.39)} & 32.1 (3.55) & 6.4 (0.55) & \bf{2.1 (0.12)} \\
NeRDm & 256 & 90.5 (0.16) & 82.8 (0.25) & \bf{29.6 (2.62)} & \bf{6.3 (0.03)} & \bf{2.1 (0.12)} \\
\hline
U-Net & 512 & 90.1 (0.14) & 82.2 (0.24) & 34.3 (0.36) & 6.9 (0.10) & 2.3 (0.08) \\
NeRDc & 512 & \bf{90.7 (0.14)} & \bf{83.1 (0.23)} & \bf{30.6 (0.67)} & \bf{6.3 (0.21)} & \bf{2.1 (0.07)}\\
NeRDm & 512 & 90.3 (0.21) & 82.5 (0.37) & 35.2 (2.49) & 6.5 (0.31) & 2.2 (0.06) \\
\hline
\hline
\end{tabular}
\end{center}
\end{table}

\begin{figure}[!t]
\centering
\includegraphics[width=1.0\columnwidth,height=0.3123\columnwidth]{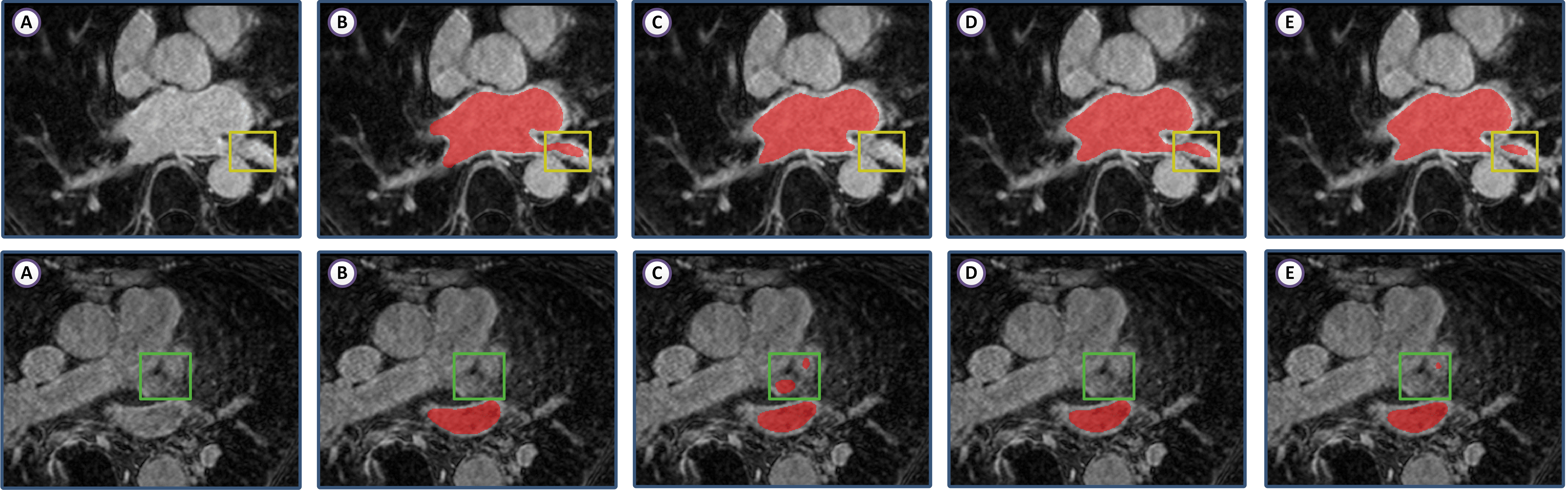}  
\caption{
LA examples. 
Yellow boxes indicate regions of missing and green boxes indicate regions of over-segmenting. 
(A) LGE-MR images; (B) ground-truth (marked in red); (C) U-Net mask; (D) NeRDc mask; (E) NeRDm mask.
}
\label{fig:atrial_example}
\end{figure}

\subsubsection{Quantitative results.}
We report the LA segmentation results of the baseline U-Net~\cite{ronneberger2015u}, NeRDc, and NeRDm in Table~\ref{tab:la_table}.
We evaluated the performance of each model using a set of boundary-based metrics, such as Hausdorff distance (HD) and average surface distance (ASD).
Similar to WMH segmentation, we employ CNN backbone with both low- and high-capacity for detailed performance investigation.
In the low-capacity group, we can observe that both NeRDc and NeRDm show improved performance on all metrics compared to U-Net.
Especially, NeRDm achieved 4.5 reduction in HD with reduced variance.
On the other hand, NeRDc outperforms both U-Net and NeRDm on all metrics in the high-capacity group with a significant margin in HD.
Notably, NeRDc in the low-capacity group showed substantial improvement over U-Net in the high-capacity group on all metrics.
This phenomena were observed in both WMH and LA segmentation, which indicates that even with fewer network parameters and lower computational resource requirement, our proposed NeRD can achieve comparable or better performance than the counterpart without NeRD module consistently.

\subsubsection{Qualitative results.}
Fig.~\ref{fig:atrial_example} shows two examples of LA segmentation results.
As can be seen in the first row of the figure, similar to WHM segmentation, U-Net equipped with the NeRD module didn't miss the pixels close to the boundary, while the U-Net did.
Similarly, our proposed methods didn't over-segment the areas close to the center of the image, while the U-Net did.
\section{Conclusions}

We presented a novel neural representation learning technique (NeRD) to estimate the pixel-wise feature distribution.
We instantiated two variants of the NeRD, NeRDm for estimating the mean and variance of the feature distribution and NeRDc for estimating pixel-wise linear classifier.
Both variants showed performance improvement over the counterpart without NeRD modules on two challenging medical image segmentation tasks, WHM segmentation that contains multiple lesions spanning the whole brain and LA segmentation that contains a single large object.
We believe that the proposed NeRD technique can contribute to more medical image applications.


\bibliographystyle{splncs04}
\bibliography{mybibliography}

\begin{thebibliography}{10}
\providecommand{\url}[1]{\texttt{#1}}
\providecommand{\urlprefix}{URL }
\providecommand{\doi}[1]{https://doi.org/#1}

\bibitem{alsallakh2020mind}
Alsallakh, B., Kokhlikyan, N., Miglani, V., Yuan, J., Reblitz-Richardson, O.:
  Mind the pad--cnns can develop blind spots. arXiv preprint arXiv:2010.02178
  (2020)

\bibitem{anokhin2020image}
Anokhin, I., Demochkin, K., Khakhulin, T., Sterkin, G., Lempitsky, V.,
  Korzhenkov, D.: Image generators with conditionally-independent pixel
  synthesis. arXiv preprint arXiv:2011.13775  (2020)

\bibitem{bengio2013better}
Bengio, Y., Mesnil, G., Dauphin, Y., Rifai, S.: Better mixing via deep
  representations. In: International conference on machine learning. pp.
  552--560. PMLR (2013)

\bibitem{chen2020learning}
Chen, Y., Liu, S., Wang, X.: Learning continuous image representation with
  local implicit image function. arXiv preprint arXiv:2012.09161  (2020)

\bibitem{cheng2018cube}
Cheng, H.T., Chao, C.H., Dong, J.D., Wen, H.K., Liu, T.L., Sun, M.: Cube
  padding for weakly-supervised saliency prediction in 360 videos. In:
  Proceedings of the IEEE Conference on Computer Vision and Pattern
  Recognition. pp. 1420--1429 (2018)

\bibitem{dice1945measures}
Dice, L.R.: Measures of the amount of ecologic association between species.
  Ecology  \textbf{26}(3),  297--302 (1945)

\bibitem{hu2018learning}
Hu, R., Doll{\'a}r, P., He, K., Darrell, T., Girshick, R.: Learning to segment
  every thing. In: Proceedings of the IEEE Conference on Computer Vision and
  Pattern Recognition. pp. 4233--4241 (2018)

\bibitem{hu2019meta}
Hu, X., Mu, H., Zhang, X., Wang, Z., Tan, T., Sun, J.: Meta-sr: A
  magnification-arbitrary network for super-resolution. In: Proceedings of the
  IEEE Conference on Computer Vision and Pattern Recognition. pp. 1575--1584
  (2019)

\bibitem{innamorati2019learning}
Innamorati, C., Ritschel, T., Weyrich, T., Mitra, N.J.: Learning on the edge:
  Investigating boundary filters in cnns. International Journal of Computer
  Vision pp. 1--10 (2019)

\bibitem{islam2019much}
Islam, M.A., Jia, S., Bruce, N.D.: How much position information do
  convolutional neural networks encode? In: International Conference on
  Learning Representations (2019)

\bibitem{jiang2020local}
Jiang, C., Sud, A., Makadia, A., Huang, J., Nie{\ss}ner, M., Funkhouser, T.,
  et~al.: Local implicit grid representations for 3d scenes. In: Proceedings of
  the IEEE/CVF Conference on Computer Vision and Pattern Recognition. pp.
  6001--6010 (2020)

\bibitem{kayhan2020translation}
Kayhan, O.S., Gemert, J.C.v.: On translation invariance in cnns: Convolutional
  layers can exploit absolute spatial location. In: Proceedings of the IEEE/CVF
  Conference on Computer Vision and Pattern Recognition. pp. 14274--14285
  (2020)

\bibitem{kervadec2019boundary}
Kervadec, H., Bouchtiba, J., Desrosiers, C., Granger, E., Dolz, J., Ayed, I.B.:
  Boundary loss for highly unbalanced segmentation. In: International
  conference on medical imaging with deep learning. pp. 285--296. PMLR (2019)

\bibitem{kingma2014adam}
Kingma, D.P., Ba, J.: Adam: A method for stochastic optimization. arXiv
  preprint arXiv:1412.6980  (2014)

\bibitem{kuijf2019standardized}
Kuijf, H.J., Biesbroek, J.M., De~Bresser, J., Heinen, R., Andermatt, S., Bento,
  M., Berseth, M., Belyaev, M., Cardoso, M.J., Casamitjana, A., et~al.:
  Standardized assessment of automatic segmentation of white matter
  hyperintensities and results of the wmh segmentation challenge. IEEE
  transactions on medical imaging  \textbf{38}(11),  2556--2568 (2019)

\bibitem{la2020multiple}
La~Rosa, F., Abdulkadir, A., Fartaria, M.J., Rahmanzadeh, R., Lu, P.J.,
  Galbusera, R., Barakovic, M., Thiran, J.P., Granziera, C., Cuadra, M.B.:
  Multiple sclerosis cortical and wm lesion segmentation at 3t mri: a deep
  learning method based on flair and mp2rage. NeuroImage: Clinical
  \textbf{27},  102335 (2020)

\bibitem{lemke2015metalearning}
Lemke, C., Budka, M., Gabrys, B.: Metalearning: a survey of trends and
  technologies. Artificial intelligence review  \textbf{44}(1),  117--130
  (2015)

\bibitem{liu2008reducing}
Liu, R., Jia, J.: Reducing boundary artifacts in image deconvolution. In: 2008
  15th IEEE International Conference on Image Processing. pp. 505--508. IEEE
  (2008)

\bibitem{mildenhall2020nerf}
Mildenhall, B., Srinivasan, P.P., Tancik, M., Barron, J.T., Ramamoorthi, R.,
  Ng, R.: Nerf: Representing scenes as neural radiance fields for view
  synthesis. arXiv preprint arXiv:2003.08934  (2020)

\bibitem{noh2016image}
Noh, H., Hongsuck~Seo, P., Han, B.: Image question answering using
  convolutional neural network with dynamic parameter prediction. In:
  Proceedings of the IEEE conference on computer vision and pattern
  recognition. pp. 30--38 (2016)

\bibitem{park2019deepsdf}
Park, J.J., Florence, P., Straub, J., Newcombe, R., Lovegrove, S.: Deepsdf:
  Learning continuous signed distance functions for shape representation. In:
  Proceedings of the IEEE Conference on Computer Vision and Pattern
  Recognition. pp. 165--174 (2019)

\bibitem{paszke2019pytorch}
Paszke, A., Gross, S., Massa, F., Lerer, A., Bradbury, J., Chanan, G., Killeen,
  T., Lin, Z., Gimelshein, N., Antiga, L., et~al.: Pytorch: An imperative
  style, high-performance deep learning library. In: Advances in Neural
  Information Processing Systems. pp. 8024--8035 (2019)

\bibitem{ronneberger2015u}
Ronneberger, O., Fischer, P., Brox, T.: U-net: Convolutional networks for
  biomedical image segmentation. In: International Conference on Medical image
  computing and computer-assisted intervention. pp. 234--241. Springer (2015)

\bibitem{schubert2019circular}
Schubert, S., Neubert, P., P{\"o}schmann, J., Pretzel, P.: Circular
  convolutional neural networks for panoramic images and laser data. In: 2019
  IEEE Intelligent Vehicles Symposium (IV). pp. 653--660. IEEE (2019)

\bibitem{upchurch2017deep}
Upchurch, P., Gardner, J., Pleiss, G., Pless, R., Snavely, N., Bala, K.,
  Weinberger, K.: Deep feature interpolation for image content changes. In:
  Proceedings of the IEEE conference on computer vision and pattern
  recognition. pp. 7064--7073 (2017)

\bibitem{xiong2021global}
Xiong, Z., Xia, Q., Hu, Z., Huang, N., Bian, C., Zheng, Y., Vesal, S.,
  Ravikumar, N., Maier, A., Yang, X., et~al.: A global benchmark of algorithms
  for segmenting the left atrium from late gadolinium-enhanced cardiac magnetic
  resonance imaging. Medical Image Analysis  \textbf{67},  101832 (2021)

\bibitem{zhang2020geometric}
Zhang, H., Zhang, J., Wang, R., Zhang, Q., Gauthier, S.A., Spincemaille, P.,
  Nguyen, T.D., Wang, Y.: Geometric loss for deep multiple sclerosis lesion
  segmentation. arXiv preprint arXiv:2009.13755  (2020)

\bibitem{zhang2020efficient}
Zhang, H., Zhang, J., Wang, R., Zhang, Q., Spincemaille, P., Nguyen, T.D.,
  Wang, Y.: Efficient folded attention for 3d medical image reconstruction and
  segmentation. arXiv preprint arXiv:2009.05576  (2020)

\bibitem{zhang2019multiple}
Zhang, H., Valcarcel, A.M., Bakshi, R., Chu, R., Bagnato, F., Shinohara, R.T.,
  Hett, K., Oguz, I.: Multiple sclerosis lesion segmentation with tiramisu and
  2.5 d stacked slices. In: International Conference on Medical Image Computing
  and Computer-Assisted Intervention. pp. 338--346. Springer (2019)

\bibitem{zhang2020fidelity}
Zhang, J., Liu, Z., Zhang, S., Zhang, H., Spincemaille, P., Nguyen, T.D.,
  Sabuncu, M.R., Wang, Y.: Fidelity imposed network edit (fine) for solving
  ill-posed image reconstruction. NeuroImage  \textbf{211},  116579 (2020)

\bibitem{zhang2020bayesian}
Zhang, J., Zhang, H., Sabuncu, M., Spincemaille, P., Nguyen, T., Wang, Y.:
  Bayesian learning of probabilistic dipole inversion for quantitative
  susceptibility mapping. In: Medical Imaging with Deep Learning. pp. 892--902.
  PMLR (2020)

\bibitem{zhang2019making}
Zhang, R.: Making convolutional networks shift-invariant again. In:
  International Conference on Machine Learning. pp. 7324--7334 (2019)

\bibitem{zill2020advanced}
Zill, D.G.: Advanced engineering mathematics. Jones \& Bartlett Publishers
  (2020)

\end{thebibliography}

\end{document}